\documentclass[sigconf, screen, timestamp, 10pt]{acmart}

\pdfoutput=1

\renewcommand\footnotetextcopyrightpermission[1]{} 
\usepackage{tikz}

\usepackage{titlesec}
\titlespacing*{\section}{0pt}{2.0\baselineskip}{1.0\baselineskip}
\titlespacing*{\subsection}{0pt}{2.0\baselineskip}{1.0\baselineskip}
\titlespacing*{\subsubsection}{0pt}{2.0\baselineskip}{1.0\baselineskip}

\begin{document}
%

\baselineskip 13pt

\title{Non--Stop \& Non--Breakable Code Review Services in a Distributed System: Detecting Issues in Real Time}
\subtitle{}



\iftrue
\author{Geunsik Lim} 
\orcid{0000-0003-1845-7132}
\email{geunsik.lim@samsung.com}
\affiliation{%
    \institution{Samsung Research}
    \city{Seoul}
    \country{Korea}}

\author{Yonghwi Kwon}
\orcid{0000-0001-8376-1240}
\email{yhwi.kwon@samsung.com}
\affiliation{%
    \institution{Samsung Research}
    \city{Seoul}
    \country{Korea}}

\author{Joonbae Park}
\orcid{0000-0002-8369-1366}
\email{joonbae.park@samsung.com}
\affiliation{%
    \institution{Samsung Research}
    \city{Seoul}
    \country{Korea}}

\author{Chul-Joo Kim}
\orcid{0000-0003-4382-7094}
\email{chuljoo1.kim@samsung.com}
\affiliation{%
    \institution{Samsung Research}
    \city{Seoul}
    \country{Korea}}

\fi

\begin{abstract}
The two most significant bottlenecks in code merging are the build process and the unit tests. However, as the number of items to be checked in a code review increases, that code review becomes a bottleneck for code merging as well. Because of the dependency structure between code review services, an error in one service affects the entire service. As a result, whenever a service error occurs, it is crucial to have methods for determining which code review service has ultimately caused the error. With the goal of achieving a non-stop \& non-breakable code review service, this paper describes an early error detection method along with a case study of the service.
\end{abstract}


\maketitle

\section{Introduction}
\label{sec:introduction}

As the complexity of and number of functions in software packages continue to increase, the code size is rapidly increasing as well. As a result, build delays and build failures of packages have a catastrophic impact on the release schedule of a software platform. In particular, build errors of packages represent the biggest obstacle in the software release process. However, as the number of software developers participating in one project repository increases, code review to improve code quality is increasingly becoming the top priority \cite{20-smartbear-cr}.

On the other hand, frequent code modifications by developers increase the cost of a code review. These problems lead to a \emph{Busy-Waiting} relationship between code development and code review. Code review productivity can be improved using code review services because they automatically perform repetitive code review items \cite{20-smartbear-cr}. However, as the internal operations of the code review service have become more complicated, and as the number of code review services has grown, a distributed system environment has been introduced wherein heterogeneous services and devices are connected and operated via a network. For this reason, the stability problem of a specific service affects the reliability of the overall service. This paper proposes a \emph{Health Check System (HCS)} that observes, diagnoses, records, and reports the health status of the service, which is the highest requirement that must be satisfied to support the non-stop \& non-breakable code review service.

The remainder of this paper is structured as follows. Section~\ref{sec:relatedwork} describes related studies. Section~\ref{sec:hcs} describes the failure detection system of decentralized services. Section~\ref{sec:evaluation} describes our case study of observation and diagnosis of services. Section~\ref{sec:conclusion} summarizes future research and conclusions.

\section{Related Work}
\label{sec:relatedwork}

This section summarizes the traditional papers examining the stability and reliability of distributed services in networks. DDI \cite{2017-access-ddi} proposes a scalable search and access method in a fully distributed P2P environment. This paper studies a semantic–-based service search framework that identifies reliable services based on context requirements. Cfar \cite{2018-chi-cfar} introduced an automated code review system to detect low--quality codes that is based on program analysis technology. Recent studies in this area have focused on the new functions of code review services or fast search methods among distributed devices. By contrast, we focus on non-stop \& non-breakable code review services that detect and address the causes of the errors occurring in distributed services.

\begin{figure}
\centering
\includegraphics[width=\linewidth]{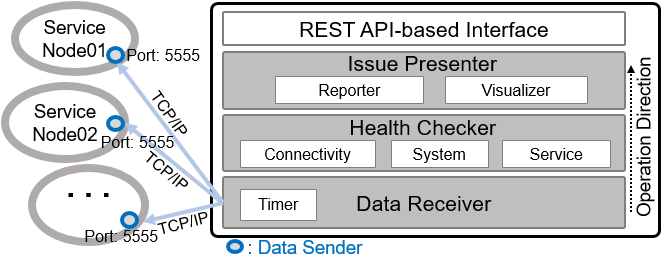}
\caption{Overall system architecture of HCS}
\label{fig:hcs}
\end{figure}

\section{HCS: Health Check System}
\label{sec:hcs}


This section describes the design and implementation of the HCS, which periodically monitors and diagnoses the health status of distributed code review services, as shown in Figure \ref{fig:hcs}.


\textbf{3.1 Data Collector:} This collects the execution information of the services distributed over the network. The \textit{Data Collector}, which operates in an N:1 network structure, primarily consists of a \textit{Data Sender} (N) that transmits the collected information and a \textit{Data Receiver} (1) to receive that information.


\textbf{3.2 Health Checker:} This is responsible for monitoring and diagnosing the health status of the code review services. The device consists of three checkers, as follows: \textit{\textbf{Connectivity Checker.}} Since all services transmit and receive information to each other over the network, the data transmission/reception speed of the services affects the speed of the code review service. For example, when services have an operation structure that is in the order of A → B → C → D, the error of B affects the reliability and stability of the entire service. Therefore, the task of checking the connectivity between services should be the top priority. \textit{\textbf{System Checker}}. To check the health status of the system in which the service is operating, the operating system-level CPU, Memory, Storage, and Network status are collected from the \textit{\/proc} filesystem. The background of this design is that system failures directly affect service disruption. \textit{\textbf{Service Checker}}. This checks the execution status of the code review services. This checker diagnoses the DB query speed of each service, the message transmission speed based on REST API, and the delay area.


\textbf{3.3 Issue Presenter:} This records the diagnostic results of the \textit{Health Checker} in the designated remote address using REST API (\textit{Reporter}), and it visualizes the observed data to improve the readability of the diagnostic results (\textit{Visualizer}). The \textit{Issue Presenter} transmits these two types of information to the developer communication channel (e.g., Slack/Mattermost or GitHub) in real time. This operation structure is intended to help service managers who need to prevent any interruptions to the code review service in advance.

\section{Evaluation}
\label{sec:evaluation}

This section describes the experimental environment used as well as a case study in which we tested our HCS to achieve the design objectives.

\begin{figure}
\centering
\includegraphics[width=\linewidth]{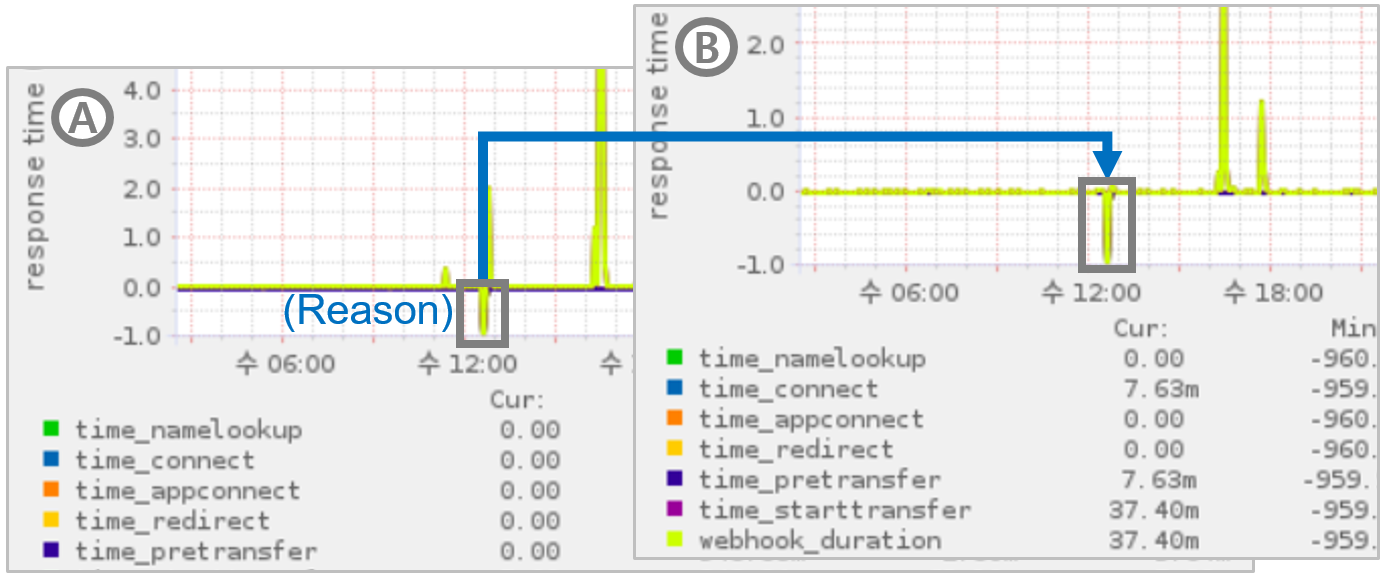}
\caption{Dependency visualization between \textcircled{\raisebox{-1.0pt}{A}} and \textcircled{\raisebox{-1.0pt}{B}} service}
\label{fig:connectivity}
\end{figure}


\begin{figure}
\centering
\includegraphics[width=\linewidth]{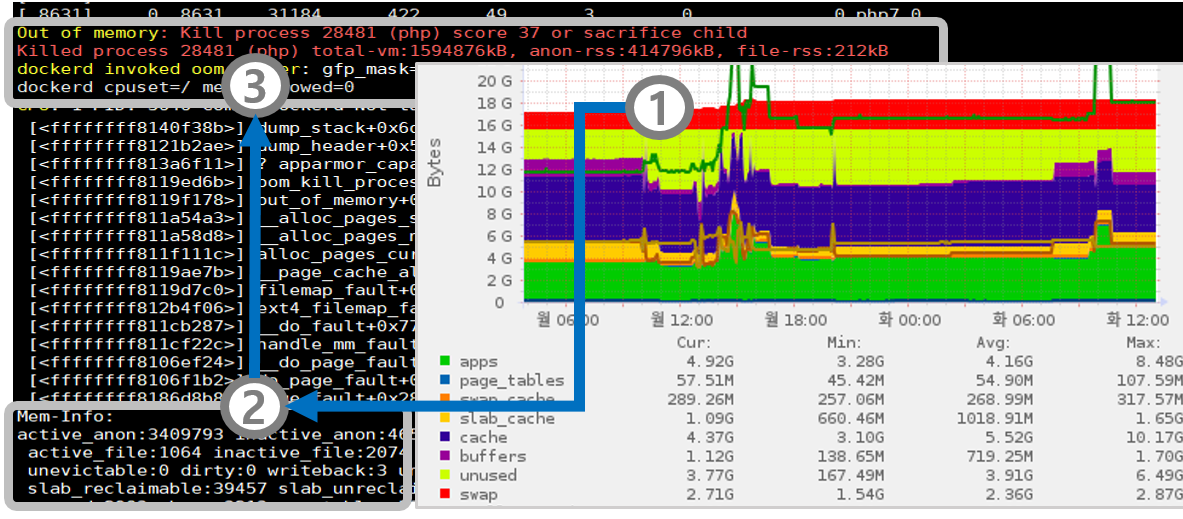}
\caption{Forced termination of services}
\label{fig:system}
\end{figure}


\begin{figure}
\centering
\includegraphics[width=\linewidth]{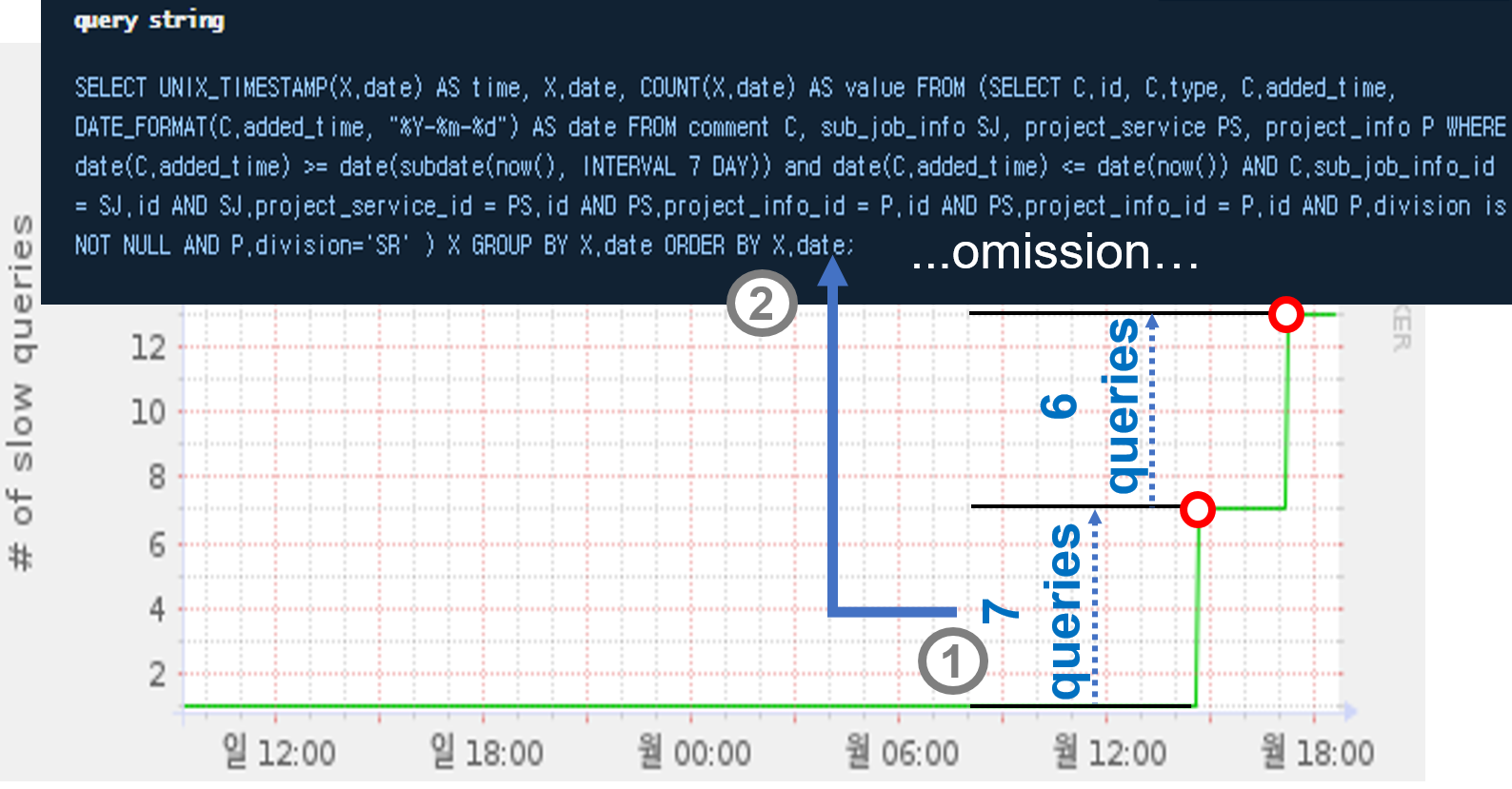}
\caption{Delay in processing speed of DB query}
\label{fig:service}
\end{figure}


\textbf{4.1 Experimental Environment:} The experimental environment is a workstation equipped with an Intel Xeon Octa-core CPU, 16 GB DDR4 RAM, 512 GB Samsung SSD, 1Gbps Ethernet, Ubuntu 16.04, Docker 19.03 (Storage Driver: overlay2), and MySQL for the code review service based on a distributed system. After the \textit{Data Receiver} collects data from the services in a designated time unit, the \textit{Health Checker} observes and diagnoses the performance degradation points and error prediction points of the services. The experimental results provide answers to the following questions. \textbf{Q1.} Do the collected information and visualization improve readability to elucidate the cause of the failure? \textbf{Q2.} Do the collected information and diagnosis results help prevent service interruptions in advance?

\textbf{4.2 Case Study:} This section describes a case study in which the proposed system diagnoses and corrects errors in code review services operating in an industrial setting. \textit{\textbf{Connectivity: Find a criminal service facility.}} \textit{Connectivity Checker} expresses the network status of each service in detail in a graph to identify code review services that cause errors in all of the services distributed over the network. Figure \ref{fig:connectivity} shows a visualization of a service dependency where \textcircled{\raisebox{-1.0pt}{C}} requires \textcircled{\raisebox{-1.0pt}{B}} and \textcircled{\raisebox{-1.0pt}{B}} requires \textcircled{\raisebox{-1.0pt}{A}}. As shown in the visualization in Figure \ref{fig:connectivity}, this process accelerates the analysis of the observation that {\em ``\textcircled{\raisebox{-1.0pt}{B}} fails because \textcircled{\raisebox{-1.0pt}{A}} fails''}. \textit{\textbf{System: Find a victim service of Out-Of-Memory Killer (OOMK).}} Figure \ref{fig:system} shows the SWAP memory usage rate (\textcircled{\raisebox{-1.0pt}{1}}) and victim process information of OOMK (\textcircled{\raisebox{-1.0pt}{2}}, \textcircled{\raisebox{-1.0pt}{3}}) provided by the System Checker when the system’s SWAP usage reaches 100\%. According to our analysis, the forced termination of the service by the operating system was the main factor hindering the uninterrupted code review service when the competition for the machine’s system resources rapidly intensified. \textit{\textbf{Service: Find a bottleneck factor from DB.}} \textit{Service Checker} diagnoses the health status of the service. Most code review services input, modify, and delete data in the relational DB. Therefore, the query speed of the DB affects the processing speed of the service. As shown in Figure \ref{fig:service}, when each DB query speed is delayed compared to the expected query speed, the \textit{Issue Presenter} records these query statements (\textcircled{\raisebox{-1.0pt}{2}}) and visualizes them in a graph (\textcircled{\raisebox{-1.0pt}{1}}).

\section{Conclusion and Future Work}
\label{sec:conclusion}


The increasing complexity of software has shifted the bottleneck of code development away from the build process and unit testing and toward code review \cite{20-smartbear-cr}. This is because, unlike in the past, many developers now collaborate on code development within a single development repository. This paper proposes an HCS for implementing the non-stop code review service in a distributed service environment comprised of heterogeneous services and devices, and it demonstrates the empirical effect through case studies. In the future, we intend to investigate non-breakable service techniques that automatically solve and report instances of performance degradation as well as operation errors of code review services distributed over the network.

\section*{Acknowledgment}
We would like to thank the anonymous reviewers, Taeksu Kim, Hyungjin Kim, and Yoonki Song, for their insightful comments and feedback.
This work was supported in part by Intelligent Dev. Assistant \& Quality Tool Development (RAJ0121ZZ-32RF), Samsung Research, Samsung Electronics Co., Ltd.

\bibliography{ref}

%
\bibliographystyle{plain}


\end{document}